\documentclass[sigconf]{acmart}

\usepackage{balance}

\usepackage[ruled,vlined,noend]{algorithm2e}
\usepackage{pdfpages}

\def\BibTeX{{\rm B\kern-.05em{\sc i\kern-.025em b}\kern-.08emT\kern-.1667em\lower.7ex\hbox{E}\kern-.125emX}}
    
\copyrightyear{2020}
\acmYear{2020}
\setcopyright{acmlicensed}
\acmConference[AIES '20] {2020 AAAI/ACM Conference on AI, Ethics, and Society}{February 7--8, 2020}{New York, NY, USA}
\acmBooktitle{2020 AAAI/ACM Conference on AI, Ethics, and Society (AIES'20), February 7--8, 2020, New York, NY, USA}
\acmPrice{15.00}
\acmDOI{10.1145/3375627.3375875}
\acmISBN{978-1-4503-7110-0/20/02}

\begin{document}

\fancyhead{}

\title{Investigating the Impact of Inclusion in Face Recognition Training Data on Individual Face Identification}

\author{Chris Dulhanty}
\email{chris.dulhanty@uwaterloo.ca}
\affiliation{%
  \institution{University of Waterloo}
  \streetaddress{200 University Avenue West}
  \city{Waterloo}
  \state{Ontario}
  \country{Canada}
  \postcode{N2L 3G1}
}

\author{Alexander Wong}
\email{a28wong@uwaterloo.ca}
\affiliation{%
  \institution{University of Waterloo}
  \streetaddress{200 University Avenue West}
  \city{Waterloo}
  \state{Ontario}
  \country{Canada}
  \postcode{N2L 3G1}
}

\begin{abstract}
    Modern face recognition systems leverage datasets containing images of hundreds of thousands of \textit{specific} individuals' faces to train deep convolutional neural networks to learn an embedding space that maps an \textit{arbitrary} individual's face to a vector representation of their identity. The performance of a face recognition system in face verification (1:1) and face identification (1:N) tasks is directly related to the ability of an embedding space to discriminate between identities. Recently, there has been significant public scrutiny into the source and privacy implications of large-scale face recognition training datasets such as MS-Celeb-1M and MegaFace, as many people are uncomfortable with their face being used to train dual-use technologies that can enable mass surveillance. However, the impact of an individual's inclusion in training data on a derived system's ability to recognize them has not previously been studied. In this work, we audit ArcFace, a state-of-the-art, open source face recognition system, in a large-scale face identification experiment with more than one million distractor images. We find a Rank-1 face identification accuracy of 79.71\% for individuals present in the model's training data and an accuracy of 75.73\% for those not present. This modest difference in accuracy demonstrates that face recognition systems using deep learning work better for individuals they are trained on, which has serious privacy implications when one considers all major open source face recognition training datasets do not obtain informed consent from individuals during their collection.
\end{abstract}

%
%


%

%
\maketitle

\section{Introduction}
Face recognition systems using Deep Convolutional Neural Networks (DCNNs) depend on the collection of large image datasets containing thousands of sets of \textit{specific} individuals' faces for training. Using this data, DCNNs learn a set of parameters that can map an \textit{arbitrary} individual's face to a feature representation, or \textit{faceprint}, that has small intra-class and large inter-class variability. The ability of a face recognition system to distinguish between identities within this embedding space depends on the size and diversity of its training data, along with its model capacity and underlying algorithms. Face recognition systems have benefited from the enabling power of Internet in the collection of large-scale image datasets and from hardware improvements in enabling efficient training of large models. Recently, increased attention to face recognition by academia, industry and government has brought new researchers, ideas and funding to the field, leading to performance improvements on benchmark tasks Labelled Faces in the Wild (LFW) \cite{LFWTech} and MegaFace \cite{nech2017level}. Consequently, face recognition systems are now being integrated into consumer and industrial electronic devices and offered as application programming interfaces (APIs) by providers such as Amazon, Microsoft, IBM, Megvii and Kairos. However, along with improved performance has come increased public discourse on the ethics of face recognition systems and their development.

Algorithmic auditing of commercial face analysis applications has uncovered disparate performance for intersectional groups across several tasks. Poor performance for darker skinned females by commercial face analysis APIs has been reported by Buolamwini, Gebru and Raji \cite{buolamwini2018gender,Raji2019}, as has lower accuracy in face identification by commercial systems with respect to lower (darker) skin reflectance by researchers at the US Department of Homeland Security \cite{cook2019demographic}. As bias in training data begets bias in model performance, efforts to create more diverse datasets for these tasks have resulted. IBM's Diversity in Faces dataset \cite{merler2019diversity}, released in January 2019, is a direct response to this body of research. Using ten established coding schemes from scientific literature, researchers annotated one million face images in an effort to advance the study of fairness and accuracy in face recognition. However, this dataset has seen public scrutiny from a different, but equally notable perspective. A March 2019 investigation by NBC News into the origins of the dataset brought to the public conversation the issue of informed consent in large-scale academic image datasets, as IBM leveraged images from Flickr with a Creative Commons Licence without notifying content owners of their use \cite{solon_2019}.  

To rationalize the collection of large-scale image datasets without explicit consent of individuals, some computer vision researchers appeal to the non-commercial nature of their work. However, work by Harvey \textit{et al.} at MegaPixels have found that authors' stated limitations on dataset use do not translate to real-world restrictions \cite{megapixels}. In the case of Microsoft's MS-Celeb-1M dataset, authors included an explicit ``non-commercial research purpose only" clause with the dataset, which was the largest publicly-available face recognition dataset at the time. However, as the dataset has been cited in published works by the research arms of many commercial entities, findings cannot easily be isolated from improvements in product offerings. As a direct result of MegaPixel's work on the ethics, origins, and privacy implications of face recognition datasets, MS-Celeb-1M  \cite{guo2016ms}, Stanford's Brainwash dataset \cite{stewart2016end} and Duke's Multi-Target, Multi-Camera dataset \cite{ristani2016MTMC} were removed from their authors' websites in June 2019. However, in the case of MS-Celeb-1M, the data remains accessible via torrents, derived datasets and other hosts \cite{megapixels}.

In addition to issues of bias and informed consent in data collection, the general use of face recognition systems by commercial and government agencies has been raised by civil rights groups and research centers, as there is no oversight for its deployment in civil society \cite{aclu,whittaker2018ai}. For these and other reasons, multiple cities in the United States have banned the use of face recognition systems for law enforcement purposes \cite{conger_2019,wu_2019,ravani_2019}. Many people are concerned with their identify being used to train the dual-use technology that is face recognition. With reports of face recognition being used by law enforcement entities to identify protesters in London \cite{bowcott_2018} and Hong Kong \cite{mozur_2019}, and measures enacted to ban face masks in the latter location \cite{yu_2019}, there is merit in understanding the impact of one's inclusion in the training data that fuels the development of these systems.

In an effort to inform the conversation about informed consent and privacy in the domain of face recognition, we conduct experiments on a state-of-the-art system. The goal of this work is to determine the impact of an individual's inclusion in face recognition training data on a derived system's ability to recognize them. To the best of the authors' knowledge, this is the first paper to investigate this relationship.

The remainder of this paper is organized in the following manner; section two outlines ethical considerations for some decisions in the design and implementation of this work, section three provides background for the taxonomy, algorithms and data used in face recognition research, section four outlines the design of experiments used to address the research question, section five presents our results and adds discussion and the paper concludes in section six.

\section{Ethical Considerations}
\subsection{Intent} 
The intent of this work is to investigate the performance of face recognition systems with respect to inclusion in training datasets. While one interpretation of this work may be to motivate efforts to mitigate demographic bias in the development of face recognition systems, it should be noted that increasing the performance of face recognition systems in any context can increase their ability to be used for oppressive purposes. In addition, due to historical societal injustices against marginalized populations and racially-biased police practices in the United States, a disproportionate number of African Americans and Hispanics are present in mugshot databases, often used by law enforcement agencies as data sources for face recognition systems  \cite{naacp,garvie2016perpetual}. These populations are therefore poised to receive a greater burden of the effects of improved face recognition systems. We therefore position this work as informing the discussion on data privacy and consent when it comes to face recognition systems and do not advocate for technical improvements without a larger discussion on the appropriate use and legality of the technology.

\subsection{Use of MS-Celeb-1M}
As noted in the introduction, the MS-Celeb-1M dataset was removed from Microsoft's website in June 2019. In a response to a Financial Times inquiry, Microsoft stated the website was retired ``because the research challenge is over'' \cite{murgia_2019}. However, a version of this dataset with detected and aligned faces from a ``cleaned'' subset of the original images is available from the Intelligent Behaviour and Understanding Group (iBUG) at Imperial College London. The dataset was offered as training data for the ``Lightweight Face Recognition Challenge \& Workshop''\footnote{\url{https://ibug.doc.ic.ac.uk/resources/lightweight-face-recognition-challenge-workshop/}} the group organized at ICCV 2019. The group has pre-trained face recognition models available as benchmarks for the challenge, trained on this data.

As this work aims to conduct experiments in a realistic setting in order to better inform the conversation around data collection processes, the analysis of a state-of-the-art model, trained on a large dataset is necessary to gain insights that are applicable to commercial applications. We therefore have decided to use the MS-Celeb-1M dataset, through its derived version offered for the ICCV 2019 Workshop, for the limited scope of this work.

\begin{table*}[t]
\centering
\caption{Prominent open-source face recognition training datasets}
\begin{tabular}{cccccc}
\toprule
\textbf{Dataset} & \textbf{Year Released} & \textbf{\# Identities} & \textbf{\# Images} & \textbf{Informed Consent Obtained?} & \textbf{Source} \\
\midrule
CASIA WebFace & 2014 & 10,575 & 494K & No & \cite{yi2014learning} \\
CelebA & 2015 & 10,177 & 203K & No & \cite{liu2015faceattributes} \\
VGGFace & 2015 & 2,622 & 2.6M & No & \cite{BMVC2015_41} \\
MS-Celeb-1M & 2016 & 99,952 & 10.0M & No & \cite{guo2016ms} \\
UMDFaces & 2016 & 8,277 & 368K & No & \cite{bansal2017umdfaces} \\
MegaFace (Challenge 2) & 2016 & 672,057 & 4.7M & No & \cite{nech2017level} \\
VGGFace2 & 2018 & 9,131 & 3.3M & No & \cite{cao2018vggface2} \\ 
\bottomrule
\end{tabular}
\label{tab:data}
\end{table*}

\section{Background}
\subsection{Face Recognition Tasks}
Within the domain of face recognition lies two categories of tasks: \textit{face verification} and \textit{face identification} \cite{learned2016labeled}. 

In face verification, the goal is to assess if a presented image matches with the reference image of an individual, often to grant access to a physical device or location. Unlocking a smartphone with one's face provides an example of face verification; a person presents their face to a phone and it is verified against a reference image of the known owner of the device. This task is referred to as 1:1 matching, as there is only one individual that the presented face image is compared against. In order to confirm a match, a threshold of similarity must be met, which can be set by the developer of a system to meet a specific level of security. Performance of a system on face verification tasks is reported in terms of accuracy; the number of correct verifications of all verification attempts.

In face identification, a \textit{gallery} of known identities is constructed from face images of individuals in advance of testing. Subsequently, a face image of unknown identity is presented to the system as the \textit{probe}. The probe is then matched for similarity with all images in the gallery, constituting 1:N matching. If the system guarantees that the identity of the probe is within the gallery of identities, the problem is considered \textit{closed-set face identification}, otherwise it is considered \textit{open-set face identification}.

Closed-set face identification tasks are common in academic benchmarks, as galleries are carefully constructed by their authors to contain all probes. In open-set face identification, a confidence threshold must be set to reject matches that do not meet a certain level of similarity. The selection of an appropriate threshold is especially relevant in high-risk applications such as law enforcement in which false positives have significant implications.

Face identification performance is reported in terms of accuracy in returning the correct identity of a probe from the gallery, or in the open-set case, no identity if the probe does not exist in the galley. Common performance metrics include Rank-1 accuracy; of all identification attempts, the number of times the correct identity in the gallery is the most similar identity to the probe, and Rank-10 accuracy; the number of times the correct identity is in the ten most similar identities to the probe.

\subsection{Deep Face Recognition}
Rapid improvements in image classification in the ImageNet Large Scale Visual Recognition Challenge (ILSVRC) \cite{russakovsky2015imagenet} by AlexNet \cite{krizhevsky2012imagenet}, ZFNet \cite{zeiler2014visualizing}, GoogLeNet \cite{szegedy2015going} and ResNet \cite{he2016deep} from 2012 to 2015 cemented the DCNN as the standard method in computer vision research and applications. While early uses of convolutional neural networks in face verification showed preliminary success \cite{chopra2005learning,huang2012learning}, it was not until the introduction of the aforementioned network architectures that the modern era of deep face recognition was in full swing. Coupled with innovations in loss function design and access to larger image datasets, modern face recognition systems have improved state-of-the-art performance on benchmark face verification and identification tasks significantly in the past six years. For a complete survey of the development of deep face recognition systems, please refer to the review paper by Wang and Deng \cite{wang2018deep}; the following is a brief summary of major milestones.

The first system to adapt findings from ILSVRC to face recognition was Facebook's DeepFace \cite{taigman2014deepface}, published in 2014 by Taigman \textit{et al.}. The nine-layer AlexNet-based model was trained on a private dataset of 4.4M images of 4K identities and achieved state-of-the-art accuracy on face verification tasks LFW and YouTube Faces (YTF) \cite{wolf2011face}, reducing the error rate by more than 50\% on the latter task.

Following this work, Google introduced FaceNet in 2015 with a major innovation in loss function design \cite{schroff2015facenet}. While the standard \textit{softmax} loss function optimized inter-class differences, researchers found that intra-class differences remained high, problematic in the domain of face recognition. To rectify this problem, the \textit{triplet loss} was introduced to jointly minimize the Euclidean distance between an anchor example and a positive example of the same identity and maximize the distance between an anchor and negative example. Using a ZFNet-based model and a private dataset of 200M images of 8M identities, they achieved state-of-the-art performance on LFW and YTF. 

Innovations in loss functions dominated the next wave of improvements in benchmark tasks,  motivated by improving discrimination between classes by making features more separable. Wen \textit{et al.} introduced the Center Loss in 2016 \cite{wen2016discriminative}, followed by Liu \textit{et al.} with the Angular Softmax in 2017 \cite{liu2017sphereface}. The Large Margin Cosine Loss was introduced in 2018 by Wang \textit{et al.} \cite{wang2018cosface}, and in 2019, Deng \textit{et al.} incorporated the Additive Angular Margin Loss into the ArcFace model \cite{deng2019arcface}, considered state-of-the-art on multiple face recognition benchmarks when published.

\begin{figure*}[]
\begin{center}
\includegraphics[width=1.0\linewidth]{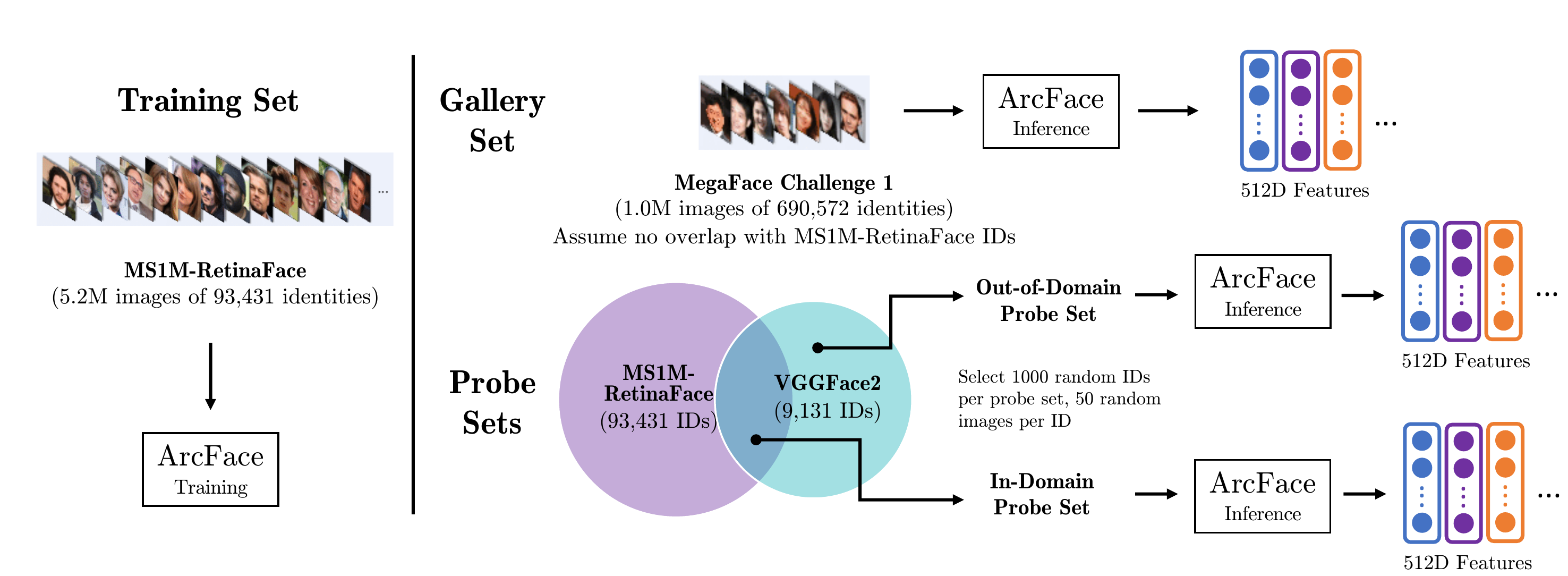}
\caption{Experimental procedure to generate feature representations of images in gallery and probe sets from ArcFace model}
\label{fig:experiments}
\end{center}
\end{figure*}

\subsection{Face Recognition Training Datasets}
Access to large-scale face recognition training datasets has been essential to the development of modern solutions by the academic community. While early published resulted in the DCNN-era of face recognition came out of companies with access to massive private datasets, such as Facebook's 500M images and 10M identities \cite{taigman2015web} and Google's 200M images and 8M identities \cite{schroff2015facenet}, the release of several open-source datasets in the ensuing years has allowed researchers to train models at scale. A summary of notable face recognition training datasets of the past six years is provided in Table \ref{tab:data}. These datasets catalyzed the field of face recognition and lead to great advances in model performance on benchmark tasks. They largely consist of celebrity identities and copyrighted images scraped from the internet.

One exception is MegaFace, which is derived from the YFCC100M dataset of 100M photos with a Creative Commons Licence, from 550K personal Flickr accounts \cite{thomee2015yfcc100m}. While the Creative Commons Licence permits the fair use of images, including in this context, Ryan Merkley, CEO of Creative Commons, noted the trouble of conflating copyright with privacy in a March 2019 statement: ``... copyright is not a good tool to protect individual privacy, to address research ethics in AI development, or to regulate the use of surveillance tools employed online. Those issues rightly belong in the public policy space, and good solutions will consider both the law and the community norms of CC licenses and content shared online in general'' \cite{cc_2019}. While MegaFace contains unknown, non-celebrity identities, an October 2019 investigation by the New York Times demonstrated that account metadata associated with images in the dataset allows for a trivial real-world identification of individuals \cite{hill_2019}. 

In all datasets, no informed consent was sought or obtained for individuals contained therein.

\section{Methodology}
\subsection{Face Recognition Model}
\subsubsection{Training Data}
We employ a cleaned version of the MS-Celeb-1M dataset \cite{guo2016ms} as training data for a face recognition model in this work. This dataset was prepared for the ICCV 2019 Lightweight Face Recognition Challenge \cite{deng2019lightweight}. All face images were preprocessed by the RetinaFace model for face detection and alignment \cite{deng2019retinaface}. A similarity transformation was applied to each detected face using five predicted face landmarks to generate normalized face crops of 112 x 112 pixels.

As the original version of this dataset has been shown to exhibit considerable inter-class noise, efforts have been made to automatically clean the dataset \cite{jin2018community}. In the case of this version, after face detection and alignment, cleaning was performed by a semi-automatic refinement strategy. First, a pre-trained ArcFace model \cite{deng2019arcface} was used to automatically remove outlier images of each identity. A manual removal of incorrectly labelled images by ``ethnicity-specific annotators" followed to result in a dataset of 5,179,510 images of 93,431 identities. We refer to this dataset as \textit{MS1M-RetinaFace}.

\subsubsection{Model}
We select the ArcFace model \cite{deng2019arcface} to study in this work. ArcFace employs the Additive Angular Margin Loss and a ResNet100 backbone to arrive at a 512-dimensional feature representation of an input image. The model achieves a verification accuracy of 99.83\% on LFW and Rank-1 identification accuracy of 81.91\% on the MegaFace Challenge 1 with one million distractors, considered state-of-the-art results. We select the model for study as is the top academic, open-source entrant on the National Institute of Standards and Technology (NIST) Face Recognition Vendor Test (FRVT) 1:1 Verification\footnote{\url{https://www.nist.gov/programs-projects/frvt-11-verification}}, a benchmark used by many commercial entities to validate the performance of their face recognition systems. Pre-trained weights for this model were provided by iBUG.

\begin{table*}[]
\centering
\caption{Face identification accuracies of ArcFace model on different probe image sets with one million distractor images}
\label{tab:res}
\begin{tabular}{ccccc}
\toprule
\textbf{Metric} & \textbf{Probe Set} & \textbf{All} & \textbf{Males} & \textbf{Females} \\
\midrule
\textbf{Rank-1 Accuracy (\%)} & In-Domain & 79.71 & 78.50 & 80.93 \\
\textbf{} & Out-of-Domain & 75.73 & 77.30 & 74.17 \\
\midrule
\textbf{Rank-10 Accuracy (\%)} & In-Domain & 90.82 & 90.92 & 90.73 \\
\textbf{} & Out-of-Domain & 86.58 & 88.59 & 84.57 \\
\midrule
\textbf{Rank-100 Accuracy (\%)} & In-Domain & 92.72 & 92.52 & 92.92 \\
\textbf{} & Out-of-Domain & 89.22 & 90.59 & 87.84 \\
\bottomrule
\end{tabular}
\end{table*}

\subsection{Experiments}
To determine the effect of inclusion in the training data of a face recognition system on its ability to identify an individual, we frame the problem as a closed-set face identification task. We construct two probe datasets and perform face identification on a gallery of one million distractor images. We assess the performance of the model on the probe datasets in terms of Rank-1, Rank-10 and Rank-100 identification accuracies. A visual representation of the datasets used in this work is shown in Figure \ref{fig:experiments}.

\subsubsection{Probe Data}
We construct two probe datasets from the VGGFace2 dataset \cite{cao2018vggface2}. Using regular expressions, we match identities in VGGFace2 by name with the identify list of MS1M-RetinaFace. We find 5,902 VGGFace2 identities present in MS1M-RetinaFace and 3,229 VGGFace2 identities not present in the training dataset. In each of these two groups, we randomly select 500 male identities and 500 female identities for evaluation, based on gender labels provided by VGGFace2 metadata. For each identity, we randomly select 50 images and perform face detection and alignment with the Multi-task Cascaded Convolutional Network (MTCNN) \cite{zhang2016joint} to generate normalized face crops of size 112 x 112 pixels. We refer to the set of 50,000 images of 1000 identities present in the training data as the \textit{in-domain probe set} and the set of 50,000 images of 1000 identities not present in the training data at the \textit{out-of-domain probe set}. We then generate 512-dimensional feature representations for all images in the in-domain and out-of-domain probe sets by running them through ArcFace.

\subsubsection{Gallery Data}
We leverage the MegaFace Challenge 1 ``Distractor'' dataset \cite{kemelmacher2016megaface} of 1,027,058 images of 690,572 identities to form the basis of the \textit{gallery}.  We again apply MTCNN to generate normalized face crops of 112 x 112 pixels for each image and run each image through ArcFace to generate 512D feature representations of all images in the gallery.

\subsubsection{Evaluation Protocol}
The experiments conducted in this work follow the protocol of MegaFace Challenge 1, with our probe sets in place of the standard FaceScrub test set  \cite{ng2014data}. We employ the Linux development kit offered by MegaFace to perform evaluation. Each probe set is evaluated following Algorithm \ref{alg}; a written description of this protocol follows.

A probe set contains 1000 identities, each with 50 images represented as 512D features. For each identity, we iterate over their images, adding one image to the gallery at a time, which we will refer to as \textit{the needle}. We then iterate over the remaining 49 images, using each one as a probe. We rank all images in the gallery by L2 distance in feature space to the probe, and record the position of the needle in the ranked list. We report results for each probe set in terms of Rank-1, Rank-10 and Rank-100 face identification accuracies.

\begin{algorithm}[]
\SetAlgoLined
\KwResult{Rank-1, 10 and 100 face identification accuracies for a probe set.}
 
 $r_1, r_{10}, r_{100} = 0$;\\
 gallery contains 1M distractor images;\\
  \For{identity in identities\textsubscript{1 to 1000}}{
    \For{image\textsubscript{needle} in images\textsubscript{1 to 50}}{
     add \textit{image\textsubscript{needle}} to the gallery;\\
     \For{image\textsubscript{probe} in images\textsubscript{1 to 50}}{
       \eIf{image\textsubscript{needle} == image\textsubscript{probe}}{
       continue;
       }{
       rank all images in gallery by L2 distance to \textit{image\textsubscript{probe}} in feature space;\\
        \If{image\textsubscript{needle} in first position in ranked list}{
         $r_1 = r_1 + 1$}
        \If{image\textsubscript{needle} in first 10 positions in ranked list}{
         $r_{10} = r_{10} + 1$}
        \If{image\textsubscript{needle} in first 100 positions in ranked list}{
         $r_{100} = r_{100} + 1$}
       }
      } remove \textit{image\textsubscript{needle}} from gallery;
     }
   }
   $\text{Rank-1\textsubscript{Acc.}} = r_1 / (1000\times50\times49)$;\\
   $\text{Rank-10\textsubscript{Acc.}} = r_{10} / (1000\times50\times49)$;\\
   $\text{Rank-100\textsubscript{Acc.}} = r_{100} / (1000\times50\times49)$;
 
 \caption{Closed-set face identification evaluation}
 \label{alg}
\end{algorithm}

\section{Results and Discussion}

We present results of the experiments in Table \ref{tab:res} for Ranks 1, 10 and 100. We find there is a modest increase in face identification accuracy for identities present in the training data, compared to those who are not. In-domain identities have a 4.0\% higher identification accuracy than out-of-domain identities at Rank-1, 4.2\% higher at Rank-10, and 3.5\% higher at Rank-100. Although not a significant margin, these results suggest that modern DCNN-based face recognition systems are biased towards individuals they are trained on.

The disparate performance between probe sets suggests some amount of overfitting has occurred in the model. Although the model generalizes well to new identities, as evidenced by results on benchmarks LFW, MegaFace and on NIST's FRVT, these results indicate that the 93k identities the system is trained on are more easily identifiable in a large-scale study. As the model's Additive Angular Margin Loss sought to increase discrimination between classes by making features more separable, it appears the model has learned to map identities to the same feature representation more consistently for those it has seen before. 

We also investigated the role of gender in the performance of the face recognition model. We find small differences in performance between genders for in-domain identities, but a 3 - 4\% decrease in performance for females compared to males who are out-of-domain, across all ranks. These results suggest that a gender bias exists in the face recognition model towards female identities. As the model has a smaller drop in face identification accuracy between domains for males, it has a greater ability to generalize to new male identities. While we do not have gender labels available for all identities in MS1M-RetinaFace, recent work has demonstrated that large-scale face recognition datasets are largely biased towards lighter-skinned males \cite{merler2019diversity}. A representational bias in MS1M-RetinaFace may account for this disparate performance across genders. Looking at these results in a different way, the consistent performance for in-domain identities across genders is perhaps more evidence that the model is overfitting to identities it has seen before. If the model only had a gender bias, we would have seen disparate performance for genders on both probe sets, however, these results suggest the model may also exhibit a ``training inclusion bias''.

Results of this study lead to the question; is the bias towards individuals in training data truly a consequence of overtraining, or is this a fundamental element of deep face recognition models? If we look to the manner by which the model was trained, overfitting in a traditional sense seems unlikely, as early stopping was employed, and results on held-out test identities demonstrate strong generalization. Perhaps there is a generalization gap in performance between in-domain and out-of-domain identities that is not apparent in current validation protocols, and increased regularization can mitigate this gap. Further testing on different training datasets and model architectures will be necessary to gather more evidence to answer this question.

We did not analyze the effect of skin type on face recognition model performance in this study, as skin type annotations were not available to us at the time.  However, two considerations were made to attempt to control for effects of skin type in these results. First, the selection of 1000 identities for each probe set is far larger than what is used in the standard protocol of MegaFace Challenge 1, where 80 identities are sampled from FaceScrub. Having a larger sample size helps to control for identities who may have either superior or poor performance due to possible model bias. In addition, the approach of random sampling in-domain and out-of-domain probe sets ensures both contain a similar distribution of identities with respect to skin type, with the assumption that the identities common to MS1M-RetinaFace and VGGFace2 and the identities distinct to VGGFace2 follow the same distribution of skin type. As both MS1M-RetinaFace and VGGFace2 use the popularity of celebrities online to construct identity lists, this assumption seems to be reasonable. Having said this, the role of skin type in the performance of the model is a very important relationship to study, and this is planned for future work. Fitzpatrick skin type \cite{fitzpatrick1988validity} annotations will need to be collected for all individuals in VGGFace2 such that sampling can be done to ensure even representation in probe sets across gender and skin type, and to determine intersectional accuracy.

The results of this study are quite concerning from a privacy and informed consent perspective. As described in the background section on Face Recognition Training Datasets, there does not exist a major open-source dataset that gathers informed consent from the individuals it contains. Without these individuals' knowledge or permission, the systems trained on their identities have a greater ability to identify them. As face recognition becomes more powerful and ubiquitous, the ability for misuse becomes greater. While MS-Celeb-1M contains only ``celebrity" identities, this classification of an individual should not negate informed consent in the development of powerful surveillance technologies. Face recognition systems are unique among biometrics as the face can be easily captured at distance without one's knowledge. The face uniquely identifies an individual, and it is difficult to opt-out of these systems without wearing a mask or other means of obfuscation, drawing undue attention to one's self. From a legal perspective, the concept of informed consent in the analysis of images of individuals' faces has traction in some jurisdictions. As reported by the New York Times with reference to potential financial liabilities of MegaFace \cite{hill_2019}, the Illinois Biometric Information Privacy Act \cite{bipa2008} is a State law enacted in 2008 that gives Illinois residents the right to seek financial compensation from entities using their face scans without their informed consent. 

The experiments in this work aim to simulate a real-world testing environment of a state-of-the-art face recognition system, with a gallery of more than one million images. These findings, therefore, may hold for systems that are currently deployed in the real-world.

\section{Conclusion}
In this work we present the first study to investigate the role of inclusion in face recognition training data on a derived system's ability to identify an individual. Through the construction of two sets of probe data that overlap and are distinct from the training data of a state-of-the-art system, we conduct a large-scale face identification experiment. We find a modest 4\% improvement in face identification accuracy for individuals who are present in training data, which is highly problematic given the norm in the field is to not gather informed consent in the collection of training datasets. Future work will apply this methodology to more models, training datasets and distance metrics (i.e. cosine distance) to see if results are consistent. Following prior work \cite{buolamwini2018gender,Raji2019,cook2019demographic}, analysis of face recognition model bias with respect to gender, skin type and their intersections in large-scale face identification tasks is needed, as well as tying results to representational bias in training data. Additionally, the relationship between the \textit{number of images} of an individual in training data and their ability to be identified is an interesting area of study. Finally, analysis of a face recognition model's feature space directly provides an alternative to a task-based auditing approach, and may be fruitful for understating nuances of inter- and intra-class differences.

%
\begin{acks}
We would like to thank the Natural Sciences and Engineering Research Council of Canada and the Canada Research Chairs Program for their support. 
\end{acks}

%
\bibliographystyle{ACM-Reference-Format}
\bibliography{AIES-bib}

\end{document}